\documentclass[prl,twocolumn,showpacs,preprintnumbers,amsmath,amssymb]{revtex4}

\usepackage{amsmath,amssymb}
\usepackage{graphicx}
\usepackage{dcolumn}
\usepackage{bm}

\begin{document}
\title{Field Compensation of Field Frequency Shifts in the Ramsey Spectroscopy of Clock Optical Transitions}
\author{A. V. Taichenachev and V. I. Yudin}
\affiliation{Institute of Laser Physics SB RAS, Novosibirsk
630090, Russia} \affiliation{Novosibirsk State University,
Novosibirsk 630090, Russia}
\author{C. W. Oates, Z. W. Barber and L. Hollberg}\affiliation{
National Institute of Standards and Technology, Boulder, CO 80305}
\date{\today}

\begin{abstract}
We develop the method of Ramsey spectroscopy with the use of an
additional field compensating the frequency shifts of clock
optical transitions. This method in combination with the method of
magnetically induced excitation of strongly forbidden transitions
$^1S_0$$\to$$^3P_0$ [A.V. Taichenachev, et al., Phys. Rev. Lett.
{\bf 96}, 083001 (2006)] allows the practically complete
compensation (down to the fractional level below 10$^{-17}$ with
respect to the unperturbed clock frequency) of field shifts of
various origin for even isotopes of alkaline-earth-like atoms
confined to an optical lattice at magic wavelength (lattice-based
atomic clocks). Apart from this, the new method can be used in the
two-photon Doppler-free Ramsey spectroscopy of other forbidden
optical transitions (for example, $^1S_0$$\to$$^1S_0$ and
$^1S_0$$\to$$^1D_2$) and other atoms. This opens perspectives for
development of principally new variants of primary optical
frequency standards based on free atoms. As a whole, the method of
the shift-compensating field is distinguished by simplicity and
high efficiency, and it can be applied for any transitions in any
atoms and ions.
\end{abstract}


\maketitle

The last few years were marked by great achievements in high-resolution spectroscopy and fundamental laser
metrology. First of all we should mention optical frequency standards on a single ion in an rf (Paul) trap
\cite{Oskay,Rosenband}, and atomic clocks based on a large number of neutral atoms confined to an optical lattice
on a magic wave length (so called lattice-based atomic clocks) \cite{Katori1,Katori2}. These titanic efforts are
directed to the creation of optical frequency standards with presently unattainable frequency uncertainty and
accuracy at the level 10$^{-17}$-10$^{-18}$. While the progress in ion standards is a natural consequence of
longstanding (more than 20 years) careful work, the experimental activity on lattice-based clocks started only
two-three years ago, when simultaneously in several laboratories super-narrow resonances have been obtained on
the strongly forbidden optical transition $^1S_0$$\to$$^3P_0$ of Sr \cite{Katori3,Ye1,Lemonde1,Ye2} and Yb
\cite{Bar06}. Even for so short period the very impressive results were achieved \cite{Ye3}, when the
metrological characteristics of the lattice-based clock is already better than for cesium fountain primary
standards. As a whole, this new direction (lattice-based atomic clocks) is now on a stage of rapid development,
which stimulates a generation of interesting physical ideas, including a new spectroscopic methods. We mention, as
an example, the method of magnetically induced spectroscopy for even isotopes of alkaline-earth-like atoms
\cite{TY06}, which has already been successfully tested for $^{174}$Yb \cite{Bar06} and $^{88}$Sr
\cite{Lemonde2}. Apart from this, there are other spectroscopic methods \cite{Zanon}, which are not realized yet
experimentally.

It should be recalled that the main idea of lattice-based clocks
\cite{Katori1} is related to the existence of such a magic wave
length of lattice field, for which the quadratic (with respect to
the field amplitude) Stark shifts of the levels of clock
transition are equal to each other. This leads to the practically
complete compensation of the frequency shift for the clock
transition. However, there are additional frequency shifts caused
by lattice field, which, despite of their relative smallness, can
have a principal significance for the definition of metrological
charcteristics of atomic clocks. First of all, it is the field
shift caused by the hyperpolarizability, which has been
investigated experimentally in \cite{Lemonde1,Bar08} and
theoretically in \cite{TY06_2}. Besides, recently we have
discovered the other shift originating from the quantization of
translational motion of atoms in a lattice with account for
contributions due to the magneto-dipole and quadrupole transitions
\cite{TY08}.

However, apart from these frequency shifts caused by the lattice
field, there exist shifts due to the presence of additional
fields. For example, in the case of odd isotopes of
alkaline-earth-like atoms a static magnetic field is used to
spectrally split the Zeeman structure of levels originating from
nonzero nuclear spin. Here both the linear and quadratic Zeeman
shifts of the clock transition frequency take place. In the case
of even isotopes (zero nuclear spin) a magnetic field is used in
the method of magnetically induced spectroscopy \cite{TY06},
leading to the quadratic shift only. Besides, for even isotopes
the AC Stark shift caused by the clock probe laser has an
essential (from metrological viewpoint) significance (see
numerical estimates in \cite{TY06}). Namely, the possibility to
eliminate such shifts (i.e. shifts caused by the additional
fields) is one of the main subjects of the present paper.

\begin{figure}[t]
\centerline{\scalebox{0.4}{\includegraphics{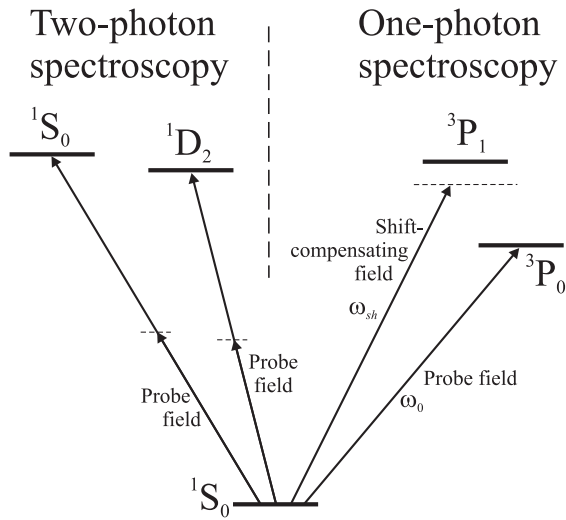}}}\caption{Scheme
of atomic levels with the indication of possible clock transitions
(to which the probe field is applied) and shift-compensating
field.}
\end{figure}

Thus, as a concrete example we will consider the method of magnetically induced spectroscopy \cite{TY06} of
strongly forbidden transitions $^1S_0$$\to$$^3P_0$ in even isotopes of alkaline-earth-like atoms (see in Fig.1)
confined to an optical lattice on the magic wavelength. For the sake of definiteness we will consider the
frequency of clock transition $\omega_0$ (i.e. in the presence only of lattice field, including the BBR shift and
collisional shift) as unperturbed. Hence we have the following frequency shift (see in \cite{TY06}):
\begin{equation}\label{shift}
\Delta=\kappa I_p +\beta|{\bf B}|^2\,,
\end{equation}
induced by the probe laser field (with the intensity $I_p$) and by the static magnetic field ${\bf B}$. Under the
creation of frequency standards (especially of primary standards) an evident question arises: what to do with
these shifts? We see two variants of answer to this question. The first variant consists in the precision
experimental measurement of the coefficients $\kappa$ and $\beta$, and the high-degree control of the values
$I_p$ and $|{\bf B}|$. Then, using results of spectroscopic measurements, one can calculate the frequency of
unperturbed transition  $\omega_0$ at $I_p$$=\,$0 and ${\bf B}$$=\,$0. For example, if we use the magnetic field
of order of $|{\bf B}|$$\sim$10$^{-3}$ T ($\sim$10 G), then the quadratic shift $\beta|{\bf B}|^2$ is of order of
10 Hz. In this case in order to have a possibility to achieve the absolute accuracy at the level of 1 mHz and
lower (the fractional error less than 10$^{-17}$) it is necessary to know the coefficient $\beta$ and to control
the value of magnetic field  $|{\bf B}|$ with the accuracy 10$^{-4}$ (for the magnetic field it means value of
order 0.1 $\mu$T). Obviously, the achievement of such a high accuracy of measurement of the coefficient $\beta$
will require a persistent efforts during many years. Probably we can not expect the same accuracy (i.e. of the
order of 0.01$\%$) of theoretical calculations for many-electron atoms due to the diamagnetic contribution.
Similar estimations can be made for the shift caused by the probe laser field ($\kappa I_p$).

\begin{figure}[t]
\centerline{\scalebox{0.4}{\includegraphics{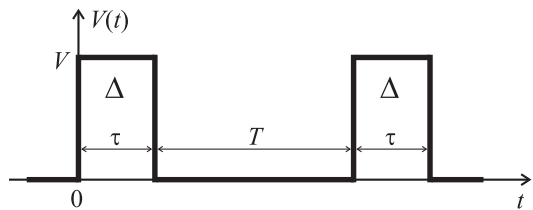}}}\caption{Illustration of the Ramsey pulses. It is shown that during the pulses the transition frequency is shifted by the value $\Delta$.}
\end{figure}

The other possible variant of solution of the problem of the field shift (\ref{shift}) is connected with the use
of the Ramsey spectroscopy with separated (in time) pulses (see in Fig.2).  As applied to the method of magnetic
field-induced spectroscopy, it is assumed here that during the pulse (with the Rabi frequency $V$ and duration
$\tau$) atoms are driven by the probe laser field with the frequency $\omega$ in the presence of static magnetic
field ${\bf B}$, and during the free evolution (of the duration $T$) both the magnetic and probe fields are
simultaneously turned off. Thus, the given version of the Ramsey spectroscopy is distinguished by the feature,
consisting in the appearance of the frequency shift $\Delta$ only during the pulse action. If in the initial
moment $t$=$\,0$ atoms were in the lower level ($^1S_0$), then after the action of two pulses the population of
atoms in the excited state ($^3P_0$) is determined by the following expression:
\begin{eqnarray}\label{n_e}
&&n_{e}=\frac{V^2}{\Omega^2}\times\\
&&\left[\cos\left(\frac{T\delta}{2}\right)\sin(2\Omega\tau)-\frac{\delta-\Delta}{\Omega}\,\sin\left(\frac{
T\delta}{2}\right)\sin^2(\Omega\tau)\right]^2,\nonumber
\end{eqnarray}
where $\delta$=($\omega$$-$$\omega_0$) is the detuning of the probe field frequency from the frequency of
unperturbed transition (i.e. during free evolution between Ramsey pulses), and
$\Omega$=$\sqrt{V^2+(\delta-\Delta)^2/4}$ is the generalized Rabi frequency.

The formula (\ref{n_e}) describes typical Ramsey fringes, where
the central resonance (as a function of $\delta$) is a reference
point for our purposes. The presence of additional frequency shift
$\Delta$ during the pulse action leads to the shift of position of
the central resonance  $\overline{\delta\omega}_0$ with respect to
the frequency of unperturbed transition $\omega_0$. If
$|\Delta/V|$$\ll\,$1, the amplitude of the central resonance is
maximal ($\approx\,$1) for $\tau V$=(2$l+$1)$\pi/4$ (where
$l$=0,1,2,...). For $l$=0 the shift of the central resonance top
$\overline{\delta\omega}_0$ (in units of $s^{-1}$) with a good
accuracy can be written as:
\begin{equation}\label{sh_res}
\overline{\delta\omega}_0\approx \xi\frac{1}{T}\,\frac{\Delta}{V}\,,
\end{equation}
where for  $1$$\leq$($VT$)$<$$\infty$ the dimensionless coefficient $\xi$ monotonically increases in the interval
0.5$\leq$$\xi$$<$1.

For Yb atoms, as it follows from the calculations in \cite{TY06}, the coefficients $\kappa$ and $\beta$ have
opposite signs, that allows one to use a specific relationship between the magnetic and probe fields ($|{\bf
B}|^2$/$I_p$=$-$$\kappa$/$\beta$), for which $\Delta$=$\,0$ (and, consequently,
$\overline{\delta\omega}_0$=$\,0$). Thus, for Yb atoms the Ramsey spectroscopy can be easily realized. However,
an attractive perspective  to use the Ramsey method for other atoms (say Sr) meets a very serious obstacle,
because the coefficients $\kappa$ and $\beta$ have the same signs. Additionally, instead of $|\Delta/V|$$\ll\,$1
the opposite condition $|\Delta/V|$$>\,$1 is usually fulfilled  for typical conditions of magnetically induced
spectroscopy \cite{TY06}. This leads to the significant shift $\overline{\delta\omega}_0$, and, obviously, in such conditions the use of Ramsey method is senseless.

Nevertheless, we find a simple way to solve this problem. It
consists in the use of an additional laser field, compensating the
frequency shift (\ref{shift}). Here any suitable transition
connected with one of the working levels can be used. For example,
one can use the laser radiation with the frequency $\omega_{sh}$
near the intercombination transition  $^1S_0$$\to$$^3P_1$ (see in
Fig.1), or near the dipole transition $^1S_0$$\to$$^1P_1$ (of
course, at sufficiently large detuning). It is assumed that the
action of the compensating field is synchronized with the Ramsey
pulse (see in Fig.2) of the probe field (with the clock transition
frequency $\omega_0$). In this case instead of the expression
(\ref{shift}) for the parameters $\Delta$ in eq.(\ref{n_e}) we
have to use the other:
\begin{equation}\label{shift2}
\Delta'=\kappa I_p +\beta|{\bf B}|^2+\eta(\omega_{sh})I_{sh}\,,
\end{equation}
where $I_{sh}$ is the intensity of the compensating field, and the
coefficient $\eta(\omega_{sh})$ governs the dependence of
additional shift on the frequency $\omega_{sh}$. In particular,
near the transition $^1S_0$$\to$$^3P_1$ the following relationship
takes place
$\eta(\omega_{sh})$$\propto\,$($\omega_{sh}-\omega_1$)$^{-1}$,
where $\omega_{1}$ is the frequency of the transition
$^1S_0$$\to$$^3P_1$ (see in Fig.1). The key idea of our method is
that the condition $\Delta'$$\approx\,$0 can be satisfied by the
proper choice of the frequency $\omega_{sh}$ and intensity
$I_{sh}$. In this case, even with account for possible variations
of the values $I_p$, ${\bf B}$, $\omega_{sh}$, and $I_{sh}$, the
regime $|\Delta'/V|$$\ll\,$1 is realized, leading to the high
efficiency of the Ramsey spectroscopy.

Let us describe the procedure of spectroscopic measurements. Initially, we fix the values $I_p$ and ${\bf B}$,
which determine the Rabi frequency $V$$\propto\,$$|{\bf B}|\sqrt{I_p}$ (see in \cite{TY06}). Then, we fix the
compensating field frequency $\omega_{sh}$, for which there certainly exists a point $I_{sh}^{(0)}$ (under
scanning of the intensity $I_{sh}$), where the condition $\Delta'$=$\,0$ is satisfied. After that measurements of
the resonance position $\overline{\delta\omega}_0$ versus the intensity $I_{sh}$ are made for several values of
the free evolution time $T$ and/or pulse duration $\tau$. The point $I_{sh}^{(0)}$ is determined as a crossing
point of these curves (see in Fig.3). And, finally, fixing the intensity $I_{sh}$=$I_{sh}^{(0)}$ and optimal
pulse duration ($\tau V$=$\pi/4$), the probe filed frequency is stabilized on the unperturbed frequency
$\omega_0$. Another way is also possible -- vice versa, to fix the intensity  $I_{sh}$ and to scan the frequency
$\omega_{sh}$ at several different values $T$ and/or $\tau$. The crossing of these curves determines the frequency
$\omega_{sh}^{(0)}$, for which $\Delta'$=$\,0$.

\begin{figure}[t]
\centerline{\scalebox{0.5}{\includegraphics{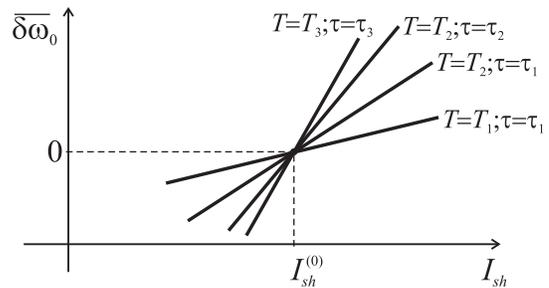}}}\caption{Illustration of the definition of the
shift-compensating field intensity $I_{sh}^{(0)}$, at which the shifts (\ref{shift2}) and (\ref{sh_res}) cancel.}
\end{figure}

The radical advantage of our method in comparison with the usual
method of single-pulse spectroscopy (currently adopted for the
lattice-based clocks) consists in that at the output we get
immediately the probe filed with the frequency $\omega_0$.
Moreover we do not need in the precision measurements (or
high-accuracy theoretical calculations) of the coefficients
$\kappa$ and $\beta$ in (\ref{shift}). Thus, the main inaccuracy
of the frequency measurements will be determined only by the
collisional shift, BBR shift, and shift due to the lattice field.
All these prove that the proposed method has essentially better
perspectives as related to the measurement of the unperturbed
frequency of clock transition. Note also that the possibility of
practical realization of the method of shift-compensating field in
the Ramsey spectroscopy does not rise any doubts in view of the
experimental realization of  the method of magnetic field-induced
spectroscopy for different atoms \cite{Bar06,Lemonde2}.

Taking Sr as an example, let us estimate the fractional frequency uncertainty of our method, using the
coefficients calculated in \cite{TY06}. Let $I_p$=100 mW/cm$^2$ and $|{\bf B}|$=0.3 mT, that leads to the values
$V$$\approx\,$0.3 Hz and $\tau$$\approx\,$0.42 $s$ ($\tau$$V$=$\pi$/4). The corresponding shift
$\Delta$$\approx$$-$3.9 Hz should be compensated by the shift-compensating field $I_{sh}$. We will assume
possible fluctuations of the intensities $I_p$ and $I_{sh}$ at level of $<$0.5$\%$, and the frequency
stabilzation at level of $<$0.5 MHz for the detuning $|\omega_{sh}-\omega_1|$$>$100 MHz. In this case for $T$=1.5
$s$ ($V$$T$$\approx\,$2.8) from the formula (\ref{sh_res}) we find the estimate of absolute frequency fluctuations
$\overline{\delta\omega}_0$$<$5 mHz, what corresponds to the fractional frequency uncertainty $<$10$^{-17}$.

The magnetic field control in our method can be substantially (in one-two order) relaxed in comparison with the
method of single-pulse magnetically induced spectroscopy \cite{TY06}. For example, if in \cite{TY06} in order to
achieve  the fractional frequency uncertainty 10$^{-17}$ for Yb it is necessary to control the magnetic field of
1 mT at level of $\sim\,$0.1 $\mu$T, then in the Ramsey method for Yb atoms it is sufficient to control the
magnetic field of 1 mT (during the pulse action) at level of 5-10 $\mu$T (including the residual field).

Let us compare now our method with the other version of the Ramsey spectroscopy proposed in \cite{Zanon} for even
isotopes of alkaline-earth atoms in lattice-based clocks. As related to the fractional frequency uncertainty, the
both methods are approximately equivalent to each other. However, in our method we always have a possibility to
achieve the maximal amplitude (100$\%$ contrast) of the central Ramsey resonance. Therefore, as a whole, the
calculated metrological characteristics of our method are better than for the method in \cite{Zanon}. As related
to technical side of the problem, then in both methods formally two lasers are used (in our method they are the
probe and shifting fields). However, in our methods they are completely independent laser sources, while in the
method of \cite{Zanon} the two lasers (with substantially different frequencies) require the strict phase locking
to each other to record the narrow resonance of the $\Lambda$ type. Thus, the experimental realization of our
method is significantly easier than in \cite{Zanon} (especially in the case of Yb atoms, for which, as it was
mentioned above, it is not necessary to use the additional shifting field). It should be also stressed that our
method is ideologically transparent and simple for theoretical analysis.

Apart from this, the proposed by us method of additional
shift-compensating field can be generalized to the case of
classical two-photon spectroscopy (in atoms and ions)
\cite{Chebotaev}, when the probe field frequency equals to the
half of the clock transition frequency. For example, the
dipole-forbidden transitions $^1S_0$$\to$$^1S_0$ or
$^1S_0$$\to$$^1D_2$ can be used, as is shown in Fig.1. Here it is
assumed that the higher levels $^1S_0$ and $^1D_2$ are long-lived
(as, for example, $^1S_0$$\to$$^1S_0$ for He, and
$^1S_0$$\to$$^1D_2$ for Ca, Ba). It is possible to work with atoms
confined to an optical lattice at the magic wave length (for the
corresponding clock transition) as well as with free atoms. Here
for free atoms it is necessary to use a field of two
counterpropagating plane waves with opposite circular
polarizations (so called $\sigma_+$-$\sigma_-$ configuration),
which allows to avoid the recoil effect and linear Doppler shift
for moving atoms \cite{Grinberg,Levenson}. Moreover, if we will
use such a field for the two-photon spectroscopy of atoms confined
to an 1D optical lattice, then its direction (the line of
wavevectors) with respect to the spatial orientation of the
lattice can be arbitrary (in particular, orthogonal). This in its
turn may be substantial technical simplification of a setup.

It should be stressed that nowadays the classical two-photon
spectroscopy is not seriously considered as a possible candidate
for the primary  frequency standards namely due to the significant
field shift $\Delta$$\propto$$I$. Indeed, in the case of usual
single-pulse spectroscopy this shift does not allow one to use the
two-photon resonance as a reference point (for the primary
frequency standards) due to fluctuations of the probe field
intensity $I$. But in the case of the Ramsey spectroscopy, apart
from the large shift, here the condition $|\Delta/V|$$>\,$1 can be
fulfilled (here $V$ is the two-photon Rabi frequency), which is
not acceptable for the effective Ramsey spectroscopy. Thus, our
method of additional shift-compensating field in the Ramsey
spectroscopy opens new perspectives for the search of other
variants of primary frequency standards. These variants can be
considered as an extension (from the viewpoint of choice of atoms
and/or clock transitions) or as a principal alternative (in the
case of free atoms) to the modern direction of lattice-based
clocks, in which the recoil effect and Doppler shift are
suppressed.

Let us add that the method of shift-compensating laser field is
quite universal, i.e. it is suitable for any transitions in any
atoms and ions. However, it can not be excluded that in some cases
the sign of the field frequency shift $\Delta$ (for the
magnetically induced spectroscopy or for the two-photon
spectroscopy) will be so that for its compensation we can use a
static electric (or even static magnetic) field, what can be a
technical simplification.

A.V.T. and V.I.Yu. were supported by RFBR (07-02-01230, 07-02-01028, 08-02-01108), INTAS-SBRAS (06-1000013-9427)
and Presidium of SB RAS.

A.V.T. and V.I.Yu. e-mail address: llf@laser.nsc.ru


\begin{thebibliography}{22}
\bibitem{Oskay} W. H. Oskay, S. A. Diddams, E. A. Donley, T. M. Fortier, T. P. Heavner, L. Hollberg,
W. M. Itano, S. R. Jefferts, M. J. Delaney, K. Kim, F. Levi, T. E. Parker, and J. C. Bergquist, Phys. Rev. Lett.
{\bf 97}, 020801 (2007).
\bibitem{Rosenband} T. Rosenband, D. B. Hume, P. O. Schmidt, et al.,
Science {\bf 319}, 1808 (2008).
\bibitem{Katori1} H. Katori, M. Takamoto, V. G. Pal'chikov, and V. D. Ovsiannikov, Phys. Rev. Lett.
{\bf 91}, 173005 (2003).
\bibitem{Katori2}
M. Takamoto and H. Katori, Phys. Rev. Lett. {\bf 91}, 223001 (2003).
\bibitem{Katori3} M. Takamoto, F.-L. Hong, R. Higashi, and H. Katori, Nature (London) {\bf 435}, 321 (2005).
\bibitem{Ye1} A. D. Ludlow, M. M. Boyd, T. Zelevinsky, S. M. Foreman, S. Blatt,
M. Notcutt, T. Ido, and J. Ye, Phys. Rev. Lett. {\bf 96}, 033003 (2006).
\bibitem{Lemonde1} A. Brusch, R. Le Targat, X. Baillard, M. Fouch\'e, and P. Lemonde, Phys. Rev. Lett.
{\bf 96}, 103003 (2006).
\bibitem{Ye2} M. M. Boyd, T. Zelevinsky, A. D. Ludlow, S. M. Foreman, S. Blatt, T. Ido, and J. Ye,
Science {\bf 314}, 1430 (2006).
\bibitem{Bar06} Z. W. Barber, C. W. Hoyt, C. W. Oates, L. Hollberg, A. V. Taichenachev, and V. I. Yudin,
Phys. Rev. Lett. {\bf 96}, 083002 (2006).
\bibitem{Ye3} A. D. Ludlow, T. Zelevinsky, G. K. Campbell, et al.,
Science {\bf 319}, 1805 (2008).
\bibitem{TY06} A.V. Taichenachev, V.I.
Yudin, C.W. Oates, C.W. Hoyt, Z.W. Barber, and L. Hollberg, Phys. Rev. Lett. {\bf 96}, 083001 (2006).
\bibitem{Lemonde2} X. Baillard, M. Fouch\'e, R. Le Targat, P.G. Westergaard, A. Lecallier, Y. Lecoq, G.D.
Rovera, S. Bize, P. Lemonde, Opt. Lett. {\bf 232}, 1812 (2007)
\bibitem{Zanon} T. Zanon-Willette, A. D. Ludlow, S. Blatt, M. M. Boyd, E. Arimondo, and J. Ye, Phys. Rev. Lett.
{\bf 97}, 233001 (2006).
\bibitem{Bar08} Z. W. Barber, J. E. Stalnaker, N. D. Lemke, N. Poli, C. W. Oates, T. M.
Fortier, S. A. Diddams, L. Hollberg, C. W. Hoyt, A. V. Taichenachev, V. I. Yudin, Phys. Rev. Lett.
{\bf 100}, 103002 (2008).
\bibitem{TY06_2} A. V. Taichenachev, V. I. Yudin, V. D. Ovsiannikov, and V. G. Pal'chikov,
Phys. Rev. Lett. {\bf 97}, 173601 (2006).
\bibitem{TY08} A. V. Taichenachev, V. I. Yudin, V. D. Ovsiannikov, and V. G. Pal'chikov,
arXiv:0803.1039 [physics.atom-ph] (2008).
\bibitem{Chebotaev} L. S. Vasilenko, V. P. Chebotaev, and A. V.
Shishaev, JETP Lett. {\bf 12}, 113 (1970).
\bibitem{Grinberg} F. Biraben, B. Cagnac, and G. Grynberg, Phys.
Rev. Lett. {\bf 32}, 643 (1974).
\bibitem{Levenson} M. D. Levenson and N. Bloembergen, Phys.
Rev. Lett. {\bf 32}, 645 (1974).

\end{thebibliography}
\end{document}